\documentclass[twocolumn,twocolappendix]{aastex63}

\usepackage{comment}
\usepackage{breqn}
\usepackage{threeparttable}

\received{September 21, 2021}
\revised{October 18, 2021}
\accepted{October 29, 2021}
\submitjournal{ApJ}

\shorttitle{Search for H$\alpha$ emitters at $z\sim7.8$}
\shortauthors{Asada \& Ohta}
\graphicspath{{./}{figures/}}
\begin{document}

\title{Search for H$\alpha$ Emitters at $z\sim7.8$: A Constraint on the H$\alpha$-based Star Formation Rate Density}

\correspondingauthor{Yoshihisa Asada}
\email{asada@kusastro.kyoto-u.ac.jp}

\author[0000-0003-3983-5438]{Yoshihisa Asada}
\affiliation{Department of Astronomy, Kyoto University \\
Sakyo-ku, Kyoto 606-8502}

\author[0000-0003-3844-1517]{Kouji Ohta}
\affiliation{Department of Astronomy, Kyoto University \\
Sakyo-ku, Kyoto 606-8502}



\begin{abstract}
We search for H$\alpha$ emitters at $z\sim7.8$ in four gravitationally lensed fields observed in the Hubble Frontier Fields program.
We use the Lyman break method to select galaxies at the target redshift, and make the photometry in {\it Spitzer}/IRAC 5.8 $\mu$m band to detect the H$\alpha$ emission from the candidate galaxies.
We find no significant detection of counterparts in the IRAC 5.8 $\mu$m band, and this gives a constraint on the H$\alpha$ luminosity function (LF) at $z\sim7.8$.
We compare the constraint with previous studies on rest-frame UV and FIR observation using the correlation between the H$\alpha$ luminosity and the star formation rate.
Additionally, we convert the constraint on the H$\alpha$ LF into an upper limit for the star formation rate density (SFRD) at this epoch assuming the shape of the LF.
We examine two types of parameterization of the LF, and obtain an upper limit for the SFRD of $\log_{10}(\rho_{\rm SFR}\ [M_\odot\ \mathrm{yr^{-1}\ Mpc^{-3}}])\lesssim-1.1$ at $z\sim7.8$.
With this constraint on the SFRD, we give an independent probe into the total star formation activity including the dust-obscured and unobscured star formation at the Epoch of Reionization.
\end{abstract}



\section{Introduction} \label{sec:intro}
One of the ultimate goals of extragalactic astronomy is the accurate quantification of the cosmic star formation history.
Clarifying how many stars formed across the cosmic time reveals the mass assembly of galaxies, which is crucial to understand the galaxy formation and evolution.
So far, a rough consensus has been reached that the cosmic star formation was most intensive at $z\sim2$-3, 2-3 billion years after the Big Bang, and got weakened to the present universe \citep[e.g.,][]{madau_cosmic_2014}.

However, the evolution of star formation rate density (SFRD) earlier than the peak at $z\sim2$-3 is still under controversy.
In this redshift range, investigation with rest UV observations has been conducted most actively.
Since UV light can be easily attenuated by dust, the correction for the loss of light due to this attenuation is essential to properly evaluate the total star formation rate (SFR).
Based on the dust-corrected rest UV observation, the total SFRD beyond $z\sim3$ is thought to be decrease as the redshift increases \citep[e.g.,][]{bouwens_alma_2020}.
SFRD measurements through FIR observations in this redshift range have also been available recently.
FIR can probe the dust-obscured star formation (SF) activity, thus comparing the results based on rest UV and FIR observations enables us to see whether the correction for the dust extinction is appropriate.
In \cite{rowan-robinson_star_2016}, the dust-obscured SF activity at $3\leq z\leq6$ is estimated to be dominant over the dust-unobscured SF, and the total SFRD is suggested to be as large as that at $z\sim2$.
Recent ALMA observations also suggest that the contribution of dust-obscured SF is significant particularly in UV-bright region even at $z\geq 7$ \citep[][]{schouws_significant_2021}.
On the other hand, in \cite{koprowski_evolving_2017}, the dust-obscured SF activity at $3\leq z\leq5$ is estimated to be small and its contribution is negligible, which is consistent with the estimation based on the dust-corrected rest UV observation.
Therefore, independent investigation is desired to measure the SFRD at $z\geq3$ more robustly.

As an independent tracer of the SF activity, the H$\alpha$ emission line is one of the most ideal indicators.
There is a tight correlation between the H$\alpha$ luminosity and the SFR among star-forming galaxies \citep[see e.g., a review by e.g.,][]{kennicutt_star_1998} and the dust extinction has much less effect on the H$\alpha$ emission than on the rest UV light.
Therefore, H$\alpha$ emission is expected to give an independent inspection on the SFRD.

At $z\geq3$, H$\alpha$ emission is redshifted and observed at $\lambda\geq 2.5\ \mu$m, so it is difficult to observe it with ground-based telescopes, and observation with a space telescope is needed.
With the current facilities, \textit{Spitzer}/IRAC is the most suitable for the observation of this wavelength range, and H$\alpha$ emission is strong enough to boost IRAC broadband photometry and thus can be detected with IRAC \citep[e.g.,][]{yabe_stellar_2009,stark_keck_2013}.
To recognize H$\alpha$ emission with broadband photometry, not only photometry at the broadband where the line falls but that at the wavelength longward of itself free from any other emission lines is also crucial.
Without an observation longward wavelength, the flux boosting by emission lines can also be interpreted as the presence of the dust reddening and/or old stellar continuum.

IRAC has four broadband filters (3.6, 4.5, 5.8 and 8.0 $\mu$m band), and the H$\alpha$ emission from $z\sim4.5,\ 5.8,\ 7.8,\ 11$ galaxies falls into each filter, respectively.
However, for $z\sim11$ galaxies, an observation at the wavelength longward of H$\alpha$ emission cannot be conducted with IRAC, and it would be difficult to measure the H$\alpha$ flux as mentioned above.
Thus, H$\alpha$ emission from $z\sim4.5$, 5.8 and 7.8 can be probed with IRAC.
Very recently, in \cite{asada_star_2021}, SFRD at $z\sim4.5$ has been estimated based on SED fitting by taking into account the effect of H$\alpha$ emission, but H$\alpha$-based SFRD at $z\sim5.8$ and 7.8 are not probed yet.
In particular, the H$\alpha$ emission from a galaxy at $z\sim7.8$ has never been reported due to its faintness.

%In investigations at these redshifts, photometries by IRAC 5.8 and/or 8.0 $\mu$m bands are crucial.
%Since the sensitivities of them are much shallower than those of 3.6 and 4.5 $\mu$m bands, only an extraordinary bright H$\alpha$ emission will be detectable.
To investigate distant and/or less luminous galaxies, gravitational lensing effect is a powerful tool.
Massive matter overdensities such as galaxy clusters can deflect the light ray from sources behind, which works just as lens does and brighten the apparent magnitude.
The Hubble Frontier Fields (HFF; P.I.: Lotz) program conducted the deepest observations towards six massive clusters with {\it Hubble} Space Telescope (HST): Abell2744, MACSJ0416, MACSJ0717, MACSJ1149, AbellS1063, and Abell370 \citep[][]{lotz_frontier_2017}, and these cluster regions are also observed with {\it Spitzer}.
A number of studies about high-$z$ galaxies have been published using the HFF data, and it has been shown that strong gravitational lensing is effective to investigate the properties of galaxies down to intrinsically faint region \citep[e.g.][]{atek_are_2015,bhatawdekar_evolution_2019,kikuchihara_early_2020}.
%Since the observation with {\it Spitzer} provides an information about rest optical emissions from high-$z$ galaxies, {\it Spitzer} data has played an important role in these studies particularly for the estimation of stellar mass \citep[e.g.][]{bhatawdekar_evolution_2019}.

In this paper, we aim at investigating the SFRD using H$\alpha$ emission line at $z\sim7.8$, which is the highest redshift that can be probed with IRAC.
The H$\alpha$ emission is probed by 5.8 $\mu m$ band with this redshift.
To investigate intrinsically faint galaxies, we utilize gravitational lensing effect.
This paper is structured as follows.
In Section \ref{sec:data}, we describe the data and make the photometry to make a sample of galaxies.
Using the photometric catalogs, we search for H$\alpha$ emitters at $z\sim7.8$, and present the resulting constraint on the H$\alpha$ luminosity function (LF) and the SFRD in Section \ref{sec:result}.
Discussions are given in Section \ref{Sec:Discussion}.
Section \ref{sec:summary} gives the summary of this paper.
Throughout this paper, all magnitudes are quoted in the AB system \citep{oke_secondary_1983}, and we use the Salpeter initial mass function.
We also assume the cosmological parameters of $H_0=70$ km s$^{-1}$ Mpc$^{-1}$, $\Omega_m = 0.3$ and $\Omega_\Lambda = 0.7$.

\section{Data}\label{sec:data}
Among the six clusters of HFF, images at 5.8 and 8.0 $\mu$m band are available for four clusters (Abell2744, MACS0717, AbellS1063, and Abell370), thus we focus on these cluster regions.
The mosaic and weight images of {\it HST} observation are created by Space Telescope Science Institute (STScI) and available on the web\footnote{\url{https://archive.stsci.edu/prepds/frontier/}}$^{,}$\footnote{\url{https://doi.org/10.17909/T9KK5N}}.
They consist of three ACS bands and four WFC3 bands: F435W ($B_{435}$), F606W ($V_{606}$), F814W ($i_{814}$), F105W ($Y_{105}$), F125W ($J_{125}$), F140W ($J\!H_{140}$), and F160W ($H_{160}$).
The mosaic and weight images of {\it Spitzer}/IRAC observation are also available\footnote{\url{https://irsa.ipac.caltech.edu/data/SPITZER/Frontier/overview.html}}.
As for the mass distribution model, we use the latest model created by The Clusters As TelescopeS (CATS) team \citep[][]{richard_mass_2014} (i.e., Version 4 model for Abell370 and Version 4.1 model for the other three clusters).

\subsection{Photometric Catalog}\label{subsec:photometry}
We use the Lyman break method to make a sample of galaxies at $6.9<z<8.6$, where the H$\alpha$ emission line falls into 5.8 $\mu$m band.
This target range of redshift can be covered by $i_{814}$- and $Y_{105}$-dropout.
Thus, we first construct photometric catalogs to obtain these Lyman Break Galaxy (LBG) candidates.
%Note that the method described below is basically identical to those is used in previous studies \citep[e.g.,][]{kawamata_precise_2016}.

To measure the {\it HST} color accurately, we homogenize the point-spread function (PSF) of {\it HST} images.
Using {\sc iraf}, we measure the sizes of PSF in {\it HST} images at $i_{814}$, $Y_{105}$, $J_{125}$, $J\!H_{140}$, and $H_{160}$ band, and convolve them with a Gaussian kernel to match the PSF sizes to the largest one.
In addition, we also make a stacked image of $B_{435}+V_{606}$.
We then run {\sc SExtractor} \citep[version 2.19.5;][]{bertin_sextractor_1996} in dual-image mode to make the photometry in each band.
The $J_{125}$ and $J\!H_{140}$ images are used as the detection image, and two photometric catalogs are constructed for each cluster field (we refer to them as $J$-selection catalog and $J\!H$-selection catalog, respectively).
We use aperture magnitudes with a diameter of $0^{\prime\prime}_{\cdot}36$ for the PSF homogenized images, and $0^{\prime\prime}_{\cdot}20$ for the non-convolved images (i.e., $B_{435}$, $V_{606}$, and $B_{435}+V_{606}$ images).
The limiting magnitudes are also measured with this diameter, which are shown in Table \ref{Tab1}.

\begin{deluxetable*}{ccccccccccccc}
\tablenum{1}\label{Tab1}
\tablecaption{$5\sigma$ Limiting Magnitudes}
\tablewidth{0pt}
\tablehead{
\colhead{Band} & \colhead{$B_{435}$} & \colhead{$V_{606}$} & \colhead{$B_{435}+V_{606}$} & \colhead{$i_{814}$\tablenotemark{a}} & \colhead{$Y_{105}$\tablenotemark{a}} & \colhead{$J_{125}$\tablenotemark{a}} & \colhead{$J\!H_{140}$\tablenotemark{a}} & \colhead{$H_{160}$\tablenotemark{a}} & \colhead{[3.6]} & \colhead{[4.5]} & \colhead{[5.8]} & \colhead{[8.0]} \\
}
\startdata
diameter & $0^{\prime\prime}_{\cdot}20$ & $0^{\prime\prime}_{\cdot}20$ & $0^{\prime\prime}_{\cdot}20$ & $0^{\prime\prime}_{\cdot}36$ & $0^{\prime\prime}_{\cdot}36$ & $0^{\prime\prime}_{\cdot}36$ & $0^{\prime\prime}_{\cdot}36$ & $0^{\prime\prime}_{\cdot}36$ & $2^{\prime\prime}_{\cdot}64$ & $2^{\prime\prime}_{\cdot}76$ & $2^{\prime\prime}_{\cdot}52$ & $2^{\prime\prime}_{\cdot}64$ \\
\hline
Abell370 & 29.61 & 29.72 & 30.28 & 28.85 & 28.91 & 28.65 & 28.75 & 28.93 & 25.05 & 24.89 & 23.13 & 24.05  \\
Abell2744 & 29.79 & 29.78 & 30.33 & 28.79 & 29.05 & 28.83 & 28.83 & 28.87 & 25.12 & 25.07 & 22.49 & 22.16  \\
MACS0717 & 29.40 & 29.76 & 30.16 & 28.72 & 28.71 & 28.50 & 28.53 & 28.36 & 24.31 & 24.47 & 20.97 & 20.93 \\
AbellS1063 & 29.47 & 29.59 & 30.36 & 28.88 & 29.10 & 28.86 & 29.03 & 28.67 & 25.10 & 24.89 & 22.14 & 22.29
\enddata
\tablenotetext{a}{Limiting magnitudes are measured for the PSF homogenized images.}
\end{deluxetable*}

\subsection{Sample Selection and IRAC photometry}\label{subsec:sample}
Using the photometric catalogs, we make a sample of candidate galaxies whose H$\alpha$ emission fall into the IRAC 5.8 $\mu$m band.
For $i_{814}$-dropout, we use $J$-selection catalog, and adopt the selection criteria which are used by \citet{ishigaki_full-data_2018} (I18, hereafter):
\begin{gather}
    i_{814} - Y_{105} > 0.8,\\
    Y_{105} - J_{125} < 0.8,\\
    i_{814} - Y_{105} > 2(Y_{105} - J_{125}) + 0.6.
\end{gather}
We require the signal-to-noise ratio (SNR) in both $Y_{105}$ and $J_{125}$ band to be larger than $5\sigma$.
In addition, we exclude any object whose SNRs in both the $B_{435}$ and $V_{606}$ band or in $B_{435}+V_{606}$ stacked image are larger than $2\sigma$.
The detection threshold in $i_{814}$ band is set to be $3\sigma$.

For $Y_{105}$-dropout, we use $J\!H$-selection catalog, and adopt the selection criteria which are used by I18 again:
\begin{gather}
    Y_{105} - J_{125} > 0.5,\\
    J_{125} - J\!H_{140} < 0.5,\\
    Y_{105} - J_{125} > 1.6(J_{125} - J\!H_{140}) + 0.4.
\end{gather}
We require the SNR in both $J_{125}$ and $J\!H_{140}$ band to be larger than $5\sigma$.
In addition, we exclude any object whose SNR in $B_{435}$, $V_{606}$, or $i_{814}$ band image is larger than $2\sigma$.
The detection threshold in $Y_{105}$ band is set to be $3\sigma$.

\begin{figure}[tpb]
\plotone{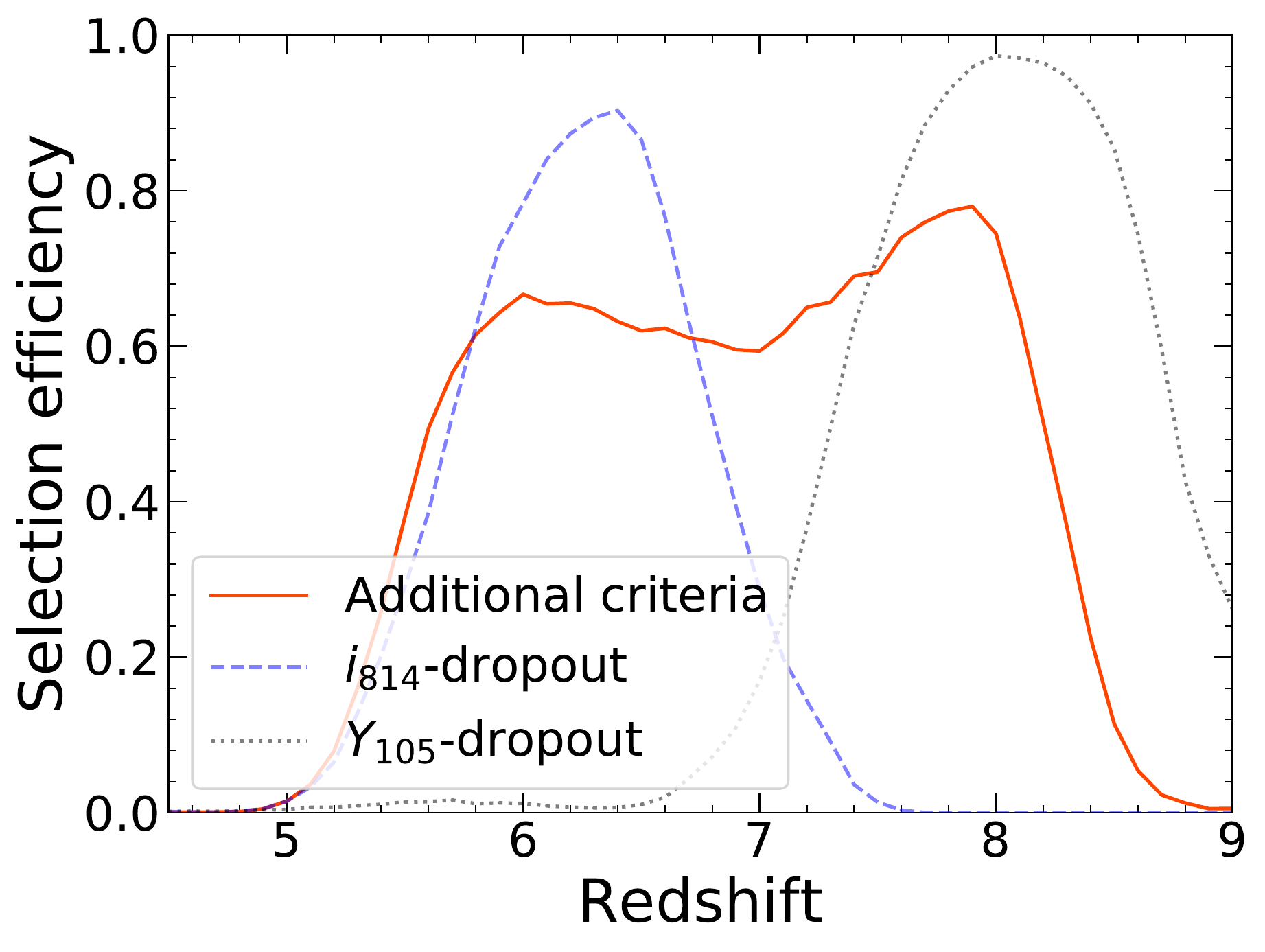}
\caption{Redshift dependence of the selection efficiency.
Blue dashed and black dotted curve shows the selection efficiency by $i_{814}$- and $Y_{105}$-dropout selection, respectively.
Both selections are insensitive to $z\sim7$ galaxies, and we use an additional selection criteria to complement our sample with them, whose selection efficiency is shown by red solid curve.
}
\label{Fig:Selection_eff}
\end{figure}

Selection efficiency against redshift is shown in Figure \ref{Fig:Selection_eff} both for $i_{814}$-dropout and $Y_{105}$-dropout selection (detail explanation for Figure \ref{Fig:Selection_eff} is given in Section \ref{subsec:Additional_cri}).
As seen in Figure \ref{Fig:Selection_eff}, these selections are insensitive for galaxies at $z\sim7$.
To make our LBG sample complemented with galaxies at $z\sim7$, we set another criteria as follows:
\begin{gather}
    i_{814}-J_{125} > 1.0, \label{eqn:7}\\
    i_{814} - J_{125} > 10(J_{125} - J\!H_{140}) - 0.2. \label{eqn:8}
\end{gather}
The criterion of equation (\ref{eqn:7}) is applied only to galaxies that are detected in $i_{814}$ band.
We require the SNR in both $Y_{105}$ and $J_{125}$ band to be larger than $5\sigma$.
In addition, we exclude any object whose SNRs in both the $B_{435}$ and $V_{606}$ band or in $B_{435}+V_{606}$ stacked image are larger than $2\sigma$.
The detection threshold in $i_{814}$ band is set to be $3\sigma$.
The selection efficiency of the additional criteria is also shown in Figure \ref{Fig:Selection_eff}.

A total of 229 galaxies are selected as our sample by the three color-color criteria.
Among the $i_{814}$-dropout or $Y_{105}$-dropout galaxies reported in the four cluster region by I18, 33 galaxies are not included in our sample of galaxies, mainly because a stacked deep image was used as the detection image in I18 while not in this study.
%In I18, 94 $i_{814}$-dropout or $Y_{105}$-dropout galaxies are detected in the four cluster region, but 33 out of the 94 galaxies are not included in our sample of galaxies, mainly because a stacked deep image was used as the detection image in I18 while not in this study.
Therefore, we add the 33 galaxies detected in I18 to our sample of galaxies.
A visual inspection is conducted to remove spurious sources for each galaxies, and about half of them are removed.
Most of the spurious sources are misidentifications of the tails of bright stars or the bad pixels at the edge of the image.
We also exclude galaxies whose redshift is spectroscopically confirmed not to be in the range of our target redshift \citep[][]{hu_redshift_2002,richard_mass_2014,vanzella_characterizing_2014,karman_muse_2015,lagattuta_lens_2017,mahler_strong-lensing_2018}.
As a result, a total of 123 galaxies are selected as our sample.
%% \footnote{A total of 134 galaxies pass the color-color selection, which include 94 galaxies that are reported as $i_{814}$-dropout or $Y_{105}$-dropout in the four cluster region by I18, and newly selected 40 galaxies.
%% For 11 galaxies out of the 134 galaxies, the redshift is confirmed not to be in the target redshift range.}

For each galaxy in the sample, we make the photometry in the {\it Spitzer}/IRAC bands.
To prioritize detection, we use aperture photometry with the optimum radius ($r_{\rm opt}\sim0.673{\rm FWHM}$) which makes the SNR best for point sources.
The center of aperture photometry is fixed to the position in the {\it HST} image that is used for detection.
The systematic difference of astrometry between {\it HST} and {\it Spitzer} image is typically less than $\sim0^{\prime\prime}_{\cdot}1$.
The limiting magnitudes for each IRAC image with this aperture photometry are made, which are also shown in Table \ref{Tab1}.

\section{Results}\label{sec:result}
\subsection{Constraint on H$\alpha$ LF}\label{subsec:constraint_HaLF}

\begin{figure*}[tpb]
\includegraphics[width=2.11\columnwidth, angle=0]{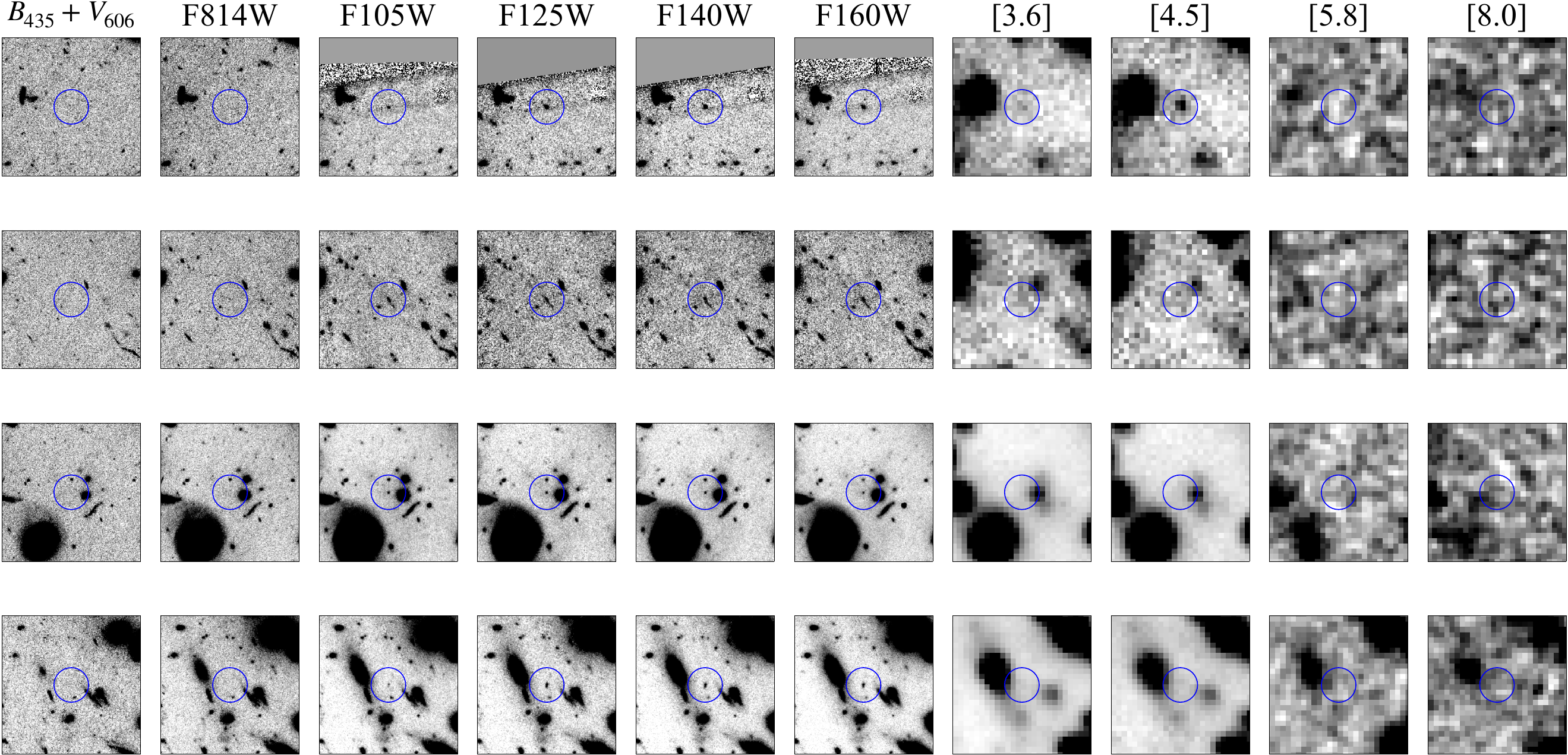}
\caption{Examples of the galaxies in the sample.
Postage stamps in the stacked image of $B_{435}+V_{606}$, ACS F814W, WFC3 F105W, F125W, F140W, F160W, IRAC 3.6 $\mu$m, 4.5 $\mu$m, 5.8 $\mu$m, 8.0 $\mu$m bands are shown for four galaxies.
The size of each cutout is $16^{\prime\prime}\times16^{\prime\prime}$, and a circle with a radius of $2^{\prime\prime}$ denotes the position of the galaxy.
}
\label{Fig:Cutouts}
\end{figure*}

We find no significant detection in IRAC 5.8 $\mu$m band at each position of the sample galaxies.
Four examples are shown in Figure \ref{Fig:Cutouts}.
Nevertheless, the H$\alpha$ LF at $z\sim7.8$ can be constrained with the fact that no H$\alpha$ emitter at $z\sim7.8$ is found by this survey.
To derive the constraint, in this section, we first derive the limiting luminosity of the H$\alpha$ emission, and put a constraint on the H$\alpha$ LF considering the gravitational lensing effect.

Using the flux density that corresponds to the detection limit of 5.8 $\mu$m band ($f^{\rm lim}_\lambda$) and the bandwidth of 5.8 $\mu$m band ($W_\lambda$), the limiting flux of 5.8 $\mu$m band can be written as $F^{\rm lim}_{\rm all} = f^{\rm lim}_\lambda W_\lambda$.
This limiting flux is attributed to the continuum ($F^{\rm lim}_{\rm cont}$) and the H$\alpha$ emission ($F^{\rm lim}_{\rm line}$), thus the following equation holds:
\begin{equation}
    F^{\rm lim}_{\rm all} = F^{\rm lim}_{\rm cont} + F^{\rm lim}_{\rm line}
\end{equation}
Considering that the maximum rest-frame equivalent width (EW) of H$\alpha$ is $4000\ {\rm \AA}$ \citep[e.g.,][]{inoue_rest-frame_2011}, we obtain
\begin{equation}
    {\rm EW}_{\rm obs} = \frac{F_{\rm line}}{f^{\rm cont}_\lambda} < 4000(1+z)\sim35000\ {\rm \AA}
\end{equation}
Since the contribution of the continuum to the limiting flux can be calculated as $F^{\rm lim}_{\rm cont} = f^{\rm lim,cont}_\lambda W_\lambda$, the ratio of contributions must obey the following inequality:
\begin{equation}
    F^{\rm lim}_{\rm line} < 35000 f^{\rm lim,cont}_\lambda = \frac{35000}{W_\lambda} F^{\rm lim}_{\rm cont}
\end{equation}
Therefore, the limiting flux of H$\alpha$ emission can be evaluated as follows:
\begin{align}
    F^{\rm lim}_{\rm line} &= \frac{F^{\rm lim}_{\rm line}}{F^{\rm lim}_{\rm line}+F^{\rm lim}_{\rm cont}} F^{\rm lim}_{\rm all} \notag \\
    &= \frac{1}{1+(F^{\rm lim}_{\rm cont}/F^{\rm lim}_{\rm line})} F^{\rm lim}_{\rm all} \notag \\
    &< \frac{1}{1+(W_\lambda/35000)}F^{\rm lim}_{\rm all} = \frac{5}{7}F^{\rm lim}_{\rm all}
\end{align}
Here, we use the value of the bandwidth of 5.8 $\mu$m band $W_\lambda = 14000\ {\rm \AA}$\footnote{This value is taken from Table 2.2 in the IRAC Instrument Handbook.}.
In the following, we use the maximum value for $F^{\rm lim}_{\rm line}$ to define the (apparent) limiting luminosity of the H$\alpha$ emission line in this observation as
\begin{equation}
    L^{\rm lim}_{\rm line} \equiv \frac{5}{7}4\pi d_L^2F^{\rm lim}_{\rm all}
\end{equation}
where $d_L$ is the luminosity distance to $z=7.8$.

\begin{figure}[tpb]
\plotone{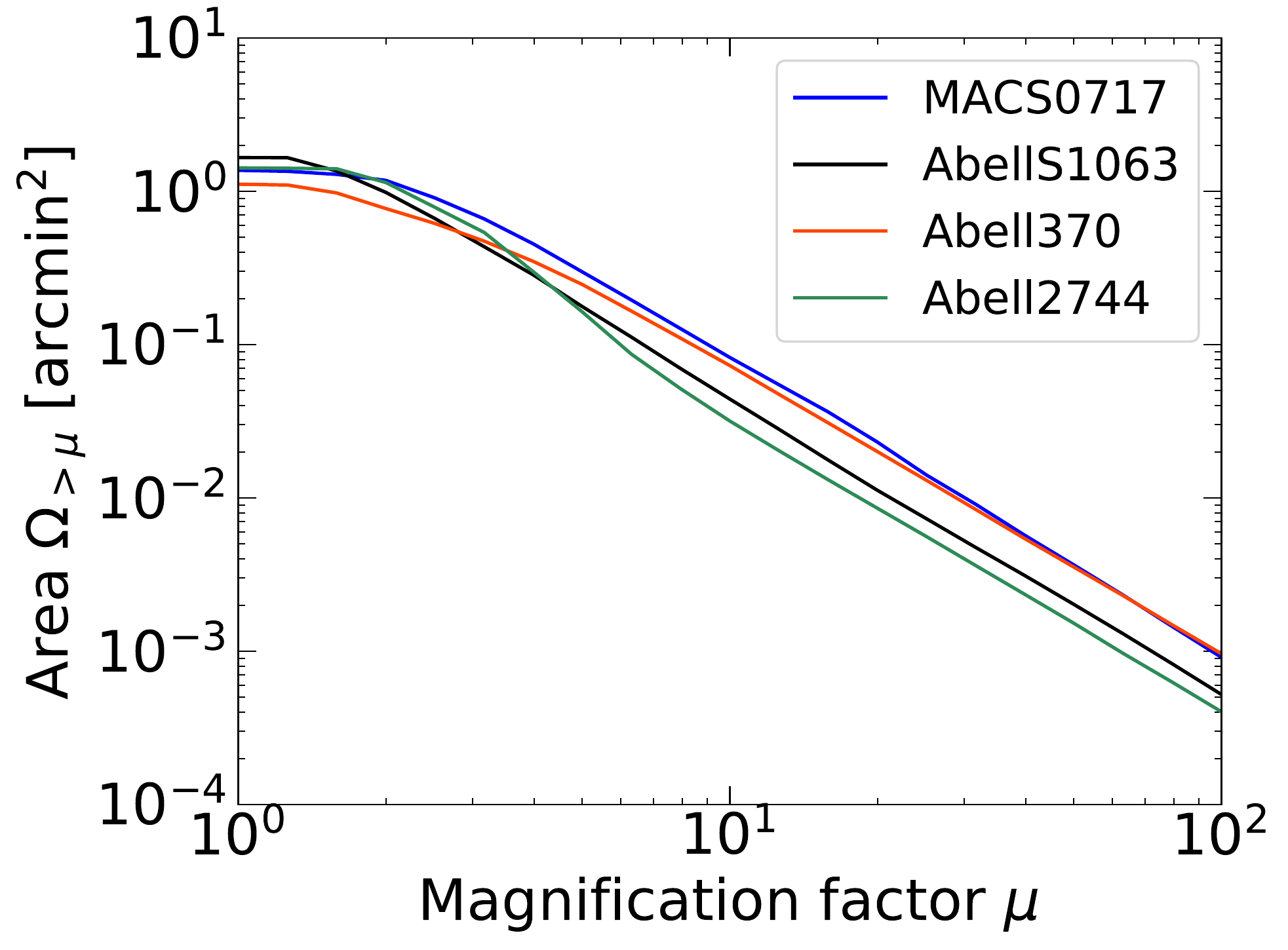}
\caption{Surface area vs. magnification factor $\mu$.
The curves represent the cumulative surface area on the source plane at $z=7.8$ for a given minimum magnification factor.
The difference of the color shows the difference of the cluster area.
The mass distribution models by CATS team are used.}
\label{Fig:Area_vs_magnif}
\end{figure}

No H$\alpha$ emission is detected with the observation whose limiting line luminosity of the H$\alpha$ emission is $L^{\rm lim}_{\rm line}$, thus the H$\alpha$ LF can be constrained.
The constraint on the LF at an intrinsic H$\alpha$ luminosity ($\phi(L^{\rm int}_{H\alpha})$) is
\begin{equation}\label{Eqn:13}
    \phi(L^{\rm int}_{H\alpha}) < \frac{1}{V_{\rm eff}(L^{\rm int}_{H\alpha})}
\end{equation}
Here, $V_{\rm eff}(L^{\rm int}_{H\alpha})$ is the effective volume where we search for galaxies with an H$\alpha$ luminosity of $L^{\rm int}_{H\alpha}$.
Using the comoving volume per unit solid angle from $z=6.9$ to $z=8.6$ (${\rm d}V$), this effective volume can be calculated as
\begin{eqnarray}\label{Eqn:14}
  V_{\rm eff}(L^{\rm int}_{H\alpha}) = \begin{cases}
  \Omega_{\mu\geq1}{\rm d}V & (L_{\rm line}^{\rm lim} < L^{\rm int}_{H\alpha})
  \\
  \Omega_{\mu\geq\mu_{\rm lim}}{\rm d}V & (L^{\rm int}_{H\alpha} < L_{\rm line}^{\rm lim})
\end{cases}
\end{eqnarray}
where $\Omega_{\mu\geq x}$ is the solid angle in the source plane where the magnification factor $\mu$ is larger than $x$, and $\mu_{\rm lim}$ is the minimum magnification factor to detect the H$\alpha$ emission of $L^{\rm int}_{H\alpha}$.
Here, we assume the source plane to be at $z=7.8$, and use $2\sigma$ limiting magnitudes at 5.8 $\mu$m band.
We show the $\Omega_{\mu\geq x}$ for each cluster region in Figure \ref{Fig:Area_vs_magnif}.

\begin{figure}[tpb]
\plotone{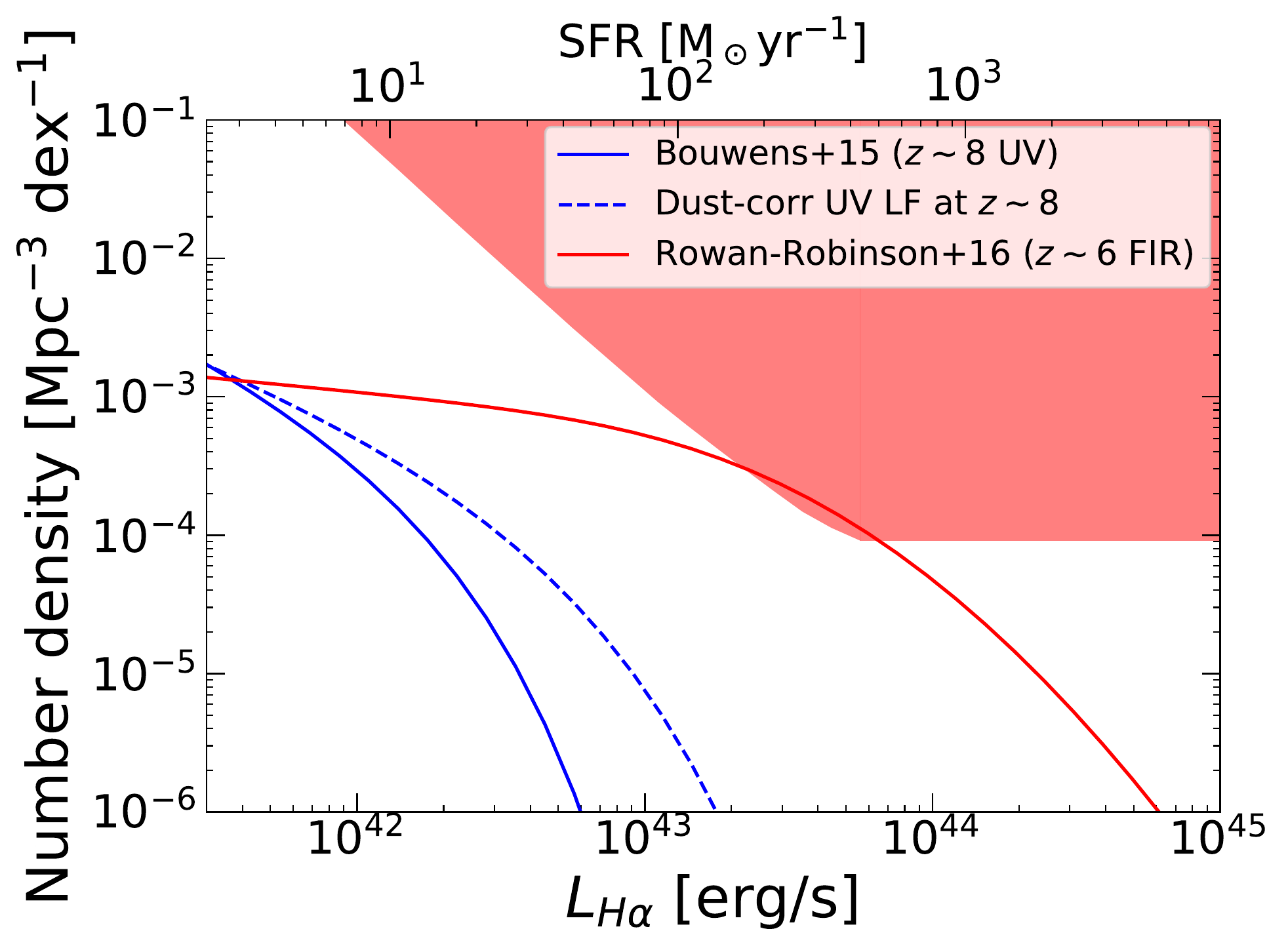}
\caption{Constraints on the H$\alpha$ LF (red shaded region) and H$\alpha$ LFs converted through the SFR from other LFs in the previous studies.
Blue solid (dashed) line shows the H$\alpha$ LF inferred from the dust-uncorrected (corrected) rest UV LF at $z\sim8$ by \citet{bouwens_uv_2015}.
Red solid line shows that inferred from the SFRF based on the FIR observation at $z\sim6$ by \citet{rowan-robinson_star_2016}.
Corresponding value of the SFR is also shown at the top ordinate.}
\label{Fig:HaLF_Constraint}
\end{figure}

The result is shown in Figure \ref{Fig:HaLF_Constraint}.
The red shaded region is ruled out by our the non-detection.
If the intrinsic H$\alpha$ luminosity is larger than the apparent limiting luminosity ($L^{\rm lim}_{\rm line}<L^{\rm int}_{{\rm H}\alpha}$), such a galaxy can be detected with this observation wherever it locates, thus the effective volume and the upper limit for the number density does not change with $L_{{\rm H}\alpha}$.
However, if the intrinsic H$\alpha$ luminosity is smaller than the apparent limiting luminosity ($L^{\rm int}_{{\rm H}\alpha}<L^{\rm lim}_{\rm line}$), such a galaxy must locate a region whose magnification factor is larger than $\mu_{\rm lim}$ to be detected with this observation.
Thus, the effective volume for such a galaxy changes with $L_{{\rm H}\alpha}$.
As can be seen in Figure \ref{Fig:Area_vs_magnif}, the larger the minimum magnification factor corresponds to the smaller effective volume.
Therefore, the upper limit for the number density of galaxies with $L_{{\rm H}\alpha}$ becomes less tight as the $L_{{\rm H}\alpha}$ is smaller.

We compare the result to the previous works on the rest UV observations and FIR observations through the SFR.
In the conversion, we use the following relations;
\begin{gather}
    L_{1500} = \left(\frac{8.0\times10^{27}}{\mathrm{erg\ s^{-1}\ Hz^{-1}}}  \right) \left(\frac{\mathrm{SFR}}{M_\odot\ \mathrm{yr}^{-1}} \right),\label{Eqn:UVtoSFR} \\
    L_{\mathrm{H}\alpha} = \left(\frac{1.3\times10^{41}}{\mathrm{erg\ s^{-1}}}  \right) \left(\frac{\mathrm{SFR}}{M_\odot\ \mathrm{yr}^{-1}} \right).\label{Eqn:HatoSFR}
\end{gather}
The equations of (\ref{Eqn:UVtoSFR}) and (\ref{Eqn:HatoSFR}) are given by \citet{madau_star_1998} and \citet{kennicutt_star_1998}, respectively.
These conversion factors can change with an assumption of different IMF (e.g., Chabrier03 IMF \citep{chabrier_galactic_2003}) or lower stellar metallicity.
However, both conversion factors changes similarly.
For example, when Chabrier03 IMF is assumed instead of Salpeter IMF, both conversion factor get larger by $\sim0.2$ dex.
When lower stellar metallicity ($Z=0.1Z_\odot$) is assumed, both conversion factor get larger by $\sim0.1$ dex.
This indicates that our result of comparison shown in Figure \ref{Fig:HaLF_Constraint} does not change even if different IMF or metallicity is assumed because the offsets on the conversion factors are canceled out.

We first compare the forbidden region to the dust-obscured SFRF based on a FIR observation at $z\sim6$ by \citet{rowan-robinson_star_2016}\footnote{The SFRF based on FIR observations is not currently probed at $z>6$.}, which is shown by red solid line in Figure \ref{Fig:HaLF_Constraint}.
If we assume that the dust-obscured SFRF does not evolve from $z\sim6$ to $z\sim7.8$\footnote{The difference of the age of the Universe at $z\sim6$ and $z\sim7.8$ is $\sim270$ Myr.}, we can see that the SFRF at $z\sim6$ is marginally consistent, or likely to overestimate the number density and/or the SFR value.
Considering that this SFRF at $z\sim6$ is derived based on extremely high SFR galaxies ($\mathrm{SFR}\sim10^4\ M_\odot\ \mathrm{yr}^{-1}$ or $L_{\mathrm{H}\alpha}\sim10^{45}\ \mathrm{erg\ s^{-1}}$) and extrapolating it down to the faint region, the result of the comparison suggests the invalidity of the extrapolation.
However, the FIR LF is obtained at $z\sim6$ and not at $z\sim7.8$, thus the result of the comparison can also suggest a negative cosmological evolution of the dust-obscured SFRF from $z\sim6$ to $z\sim8$.

To compare the forbidden region with the previous studies on rest UV observations, we correct the rest UV LF for the dust extinction following the procedure presented by \citet{smit_star_2012}.
We use the relation between the rest UV spectral slope $\beta$ and the amount of dust extinction at rest UV $A_{1600}$ given by \citet{meurer_dust_1999} and a relation between $\beta$ and (dust-uncorrected) rest UV absolute magnitude $M_{\rm UV}$ at $z\sim8$ by \citet{bouwens_uv-continuum_2014} to calculate the amount of dust extinction, and correct the (dust-uncorrected) rest UV LF at $z\sim8$ by \citet{bouwens_uv_2015}.
As can be seen in Figure \ref{Fig:HaLF_Constraint}, the dust-corrected UV LF does not violate the forbidden region, which suggests that the dust correction with this method is acceptable.

\subsection{Constraint on the SFRD}\label{subsec:constraint_SFRD}
With the non-detection of H$\alpha$ emitters at $z\sim8$, we can put a constraint on the total SFRD at this epoch.
To calculate the upper limit for the SFRD from the forbidden region for the H$\alpha$ LF, it is necessary to assume the shape of the LF.
In this work, two types of parameterization of the LF are assumed; \citet{saunders_60-micron_1990} functional form and Schechter function, which are often used in the context of FIR and rest UV observations, respectively.

\begin{comment}
\begin{figure*}[tpb]
\plotone{05_Estimate_uplims-crop.pdf}
\caption{The constraint of the total SFRD at $z\sim7.8$ based on the non-detection of H$\alpha$ emitters.
\textit{Upper left}: The upper limit for the H$\alpha$ LF at $z\sim7.8$ with the Saunders functional form.
\textit{Upper right}: The upper limit for the H$\alpha$ LF with the Schechter-like functional form.
Blue solid (dotted) line shows the dust-uncorrected (corrected) rest UV LF converted through the SFR. See the text for the details.
\textit{Bottom}: The upper limit for the total SFRD at $z\sim7.8$.
Red star (circle) shows the upper limit with Saunders (Schechter) functional form of the H$\alpha$ LF.
For clarity, these results are shifted by $\Delta z=\pm0.1$.
Black solid line shows the best-fit for the redshift evolution of the SFRD by \citet{madau_cosmic_2014}, and recent estimations using different SFR indicators are shown by open squares for comparison.
Results by \citet{rowan-robinson_star_2016}, \citet{bouwens_alma_2020}, and \citet{asada_star_2021} are based on FIR observation, rest UV observation, and SED fitting including H$\alpha$ emission, respectively.
}
\label{Fig:Uplim_for_SFRD}
\end{figure*}
\end{comment}

\begin{figure*}[tpb]
\plotone{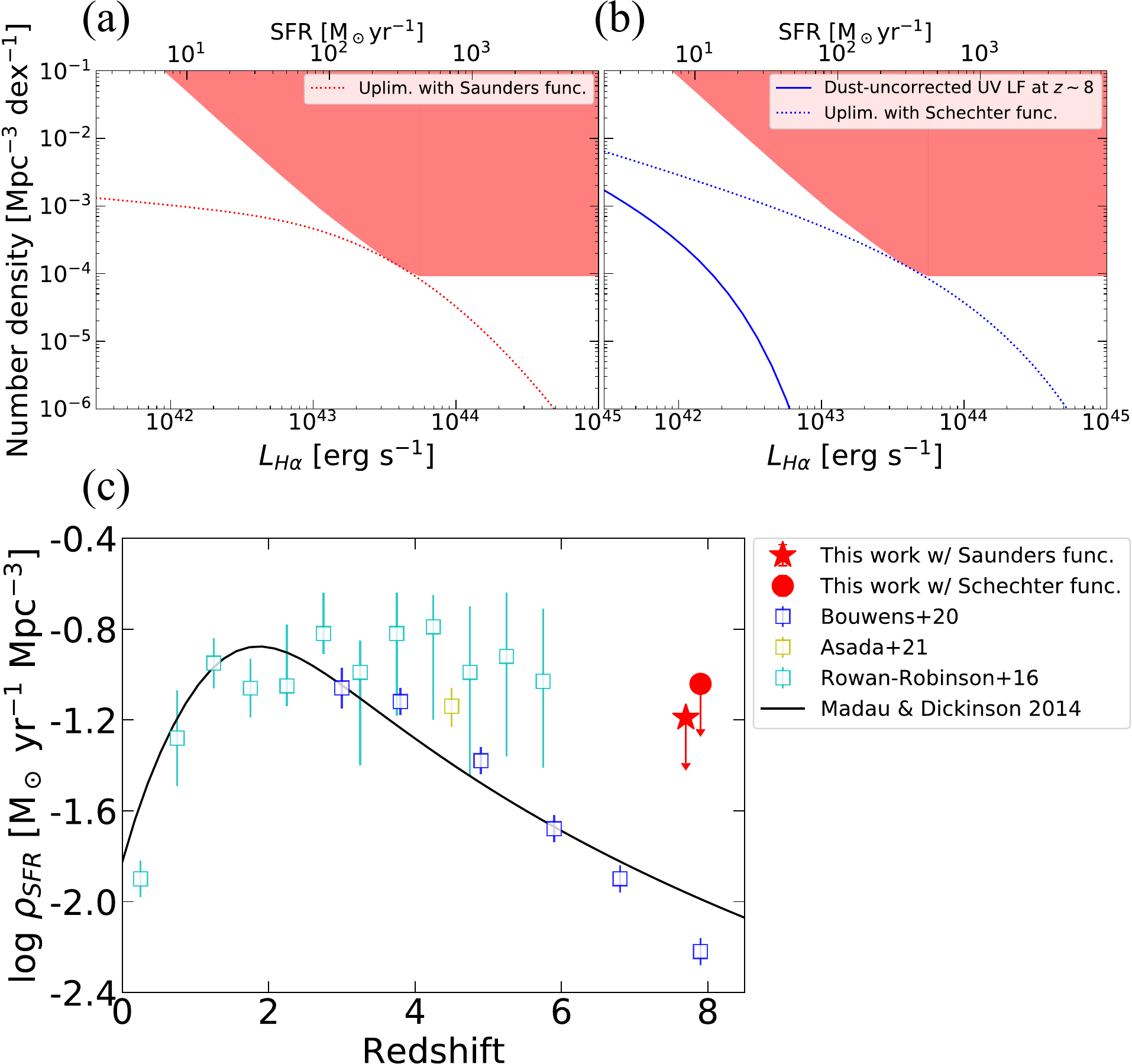}
\caption{The constraint of the total SFRD at $z\sim7.8$ based on the non-detection of H$\alpha$ emitters.
\textbf{a}, The upper limit for the H$\alpha$ LF at $z\sim7.8$ with the Saunders functional form.
\textbf{b}, The upper limit for the H$\alpha$ LF with the Schechter-like functional form.
Blue solid (dotted) line shows the dust-uncorrected (corrected) rest UV LF converted through the SFR. See the text for the details.
\textbf{c}, The upper limit for the total SFRD at $z\sim7.8$.
Red star (circle) shows the upper limit with Saunders (Schechter) functional form of the H$\alpha$ LF.
For clarity, these results are shifted by $\Delta z=\pm0.1$.
Black solid line shows the best-fit for the redshift evolution of the SFRD by \citet{madau_cosmic_2014}, and recent estimations using different SFR indicators are shown by open squares for comparison.
Results by \citet{rowan-robinson_star_2016}, \citet{bouwens_alma_2020}, and \citet{asada_star_2021} are based on FIR observation, rest UV observation, and SED fitting including H$\alpha$ emission, respectively.
}
\label{Fig:Uplim_for_SFRD}
\end{figure*}

\subsubsection{Saunders function}
To parameterize FIR LFs, a functional form that behave as power-law at low luminosity region and behave as Gaussian at high luminosity region is one of the common ways \citep[][]{saunders_60-micron_1990};
\begin{dmath}
    \phi(L)\mathrm{d}(\log L) = \ln(10)\phi^\star\left(\frac{L}{L^\star}\right)^{1-\alpha}\exp\left[ -\frac{\log^2\left(1+\frac{L}{L^\star}\right)}{2\sigma^2}\right] \mathrm{d}(\log L)
\end{dmath}
In \citet{rowan-robinson_star_2016}, this functional form is used to fit to the FIR LF, and the parameters of $\alpha$, $\sigma$, and $\phi^\star$ are found to be constant at $z>3.5$.
Thus, we fix $\alpha$, $\sigma$, and $\phi^\star$ to those by \citet{rowan-robinson_star_2016}, and examine the upper limit for $L^\star_{\mathrm{H}\alpha}$ using the forbidden region to obtain the upper limit for the SFRD with this functional form.
Consequently, we obtain the upper limit for $L^\star_{\mathrm{H}\alpha}$ to be $4.65\times10^{42}\ [\mathrm{erg\ s^{-1}}]$ (Figure \ref{Fig:Uplim_for_SFRD}(a)) and that for the SFRD to be $\log_{10}(\rho_{\rm SFR}\ [M_\odot\ \mathrm{yr^{-1}\ Mpc^{-3}}])<-1.19$.
The lower bound of the integration to calculate the SFRD is set to be $L_{\mathrm{H}\alpha}=4.44\times10^{40}\ [\mathrm{erg\ s^{-1}}]$ which corresponds to $\mathrm{SFR}\sim0.34\ [M_\odot\ \mathrm{yr^{-1}}]$ or $M_{\rm UV}\sim-17\ \mathrm{mag}$.
The result is shown in Figure \ref{Fig:Uplim_for_SFRD}(c).

\subsubsection{Schechter function}
To parameterize (dust-uncorrected) rest UV LFs, Schechter function is commonly used;
\begin{equation}
    \phi(L)\mathrm{d}(\log L) = \ln(10) \phi^\star \left(\frac{L}{L^\star}\right)^{\alpha+1}\exp\left[-\frac{L}{L^\star} \right] \mathrm{d}(\log L)
\end{equation}
To derive the upper limit for the SFRD at $z\sim8$ with this functional form, we use a (dust-uncorrected) rest UV LF at $z\sim8$ parameterized with this function, and examine the upper limit for the amount of dust extinction using the forbidden region by converting the dust-corrected UV LF to H$\alpha$ LF through the SFR.

The amount of dust extinction depends on the rest UV luminosity.
Observations and simulations reported that brighter galaxies tend to suffer from heavier dust extinction \citep[e.g.][]{smit_star_2012,yung_semi-analytic_2019,asada_star_2021}.
Thus, we assume that the correction factor $\eta$ can be written as
\begin{align}
    L_{\rm int} = \eta L,
    \\
    \eta = aL^b. \label{Eqn:eta}
\end{align}
where $L$ and $L_{\rm int}$ is the dust-uncorrected and dust-corrected rest UV luminosity\footnote{Even if we use the Meurer relation and $\beta$-$M_{\rm UV}$ relation to estimate the amount of dust extinction, the luminosity dependence of the correction factor can be expressed as equation (\ref{Eqn:eta}).}.
For simplicity, we also assume that the correction factor to be unity at the faint-end region, i.e., there is no dust extinction in the faintest galaxies, which leads to the following relation;
\begin{equation}
    1 = a\left(\frac{3\times10^{26}}{{\rm erg\ s^{-1}\ Hz^{-1}}}\right)^b \label{Eqn:Relation_ab}
\end{equation}

With these fomulations, the dust-corrected rest UV LF can be written as
\begin{dmath}
    \phi(L_{\rm int})\mathrm{d}(\log L_{\rm int}) = \ln(10) \phi_{\rm int}^\star \left(\frac{L_{\rm int}}{L_{\rm int}^\star}\right)^{\alpha_{\rm int}+1}\exp\left[-\left(\frac{L_{\rm int}}{L_{\rm int}^\star}\right)^{\frac{1}{1+b}} \right] \mathrm{d}(\log L_{\rm int})
\end{dmath}
where $\alpha_{\rm int}$, $\phi_{\rm int}$, and $L_{\rm int}^\star$ are defined using the Schechter parameters of $\alpha$, $\phi^\star$, and $L^\star$ for the dust-uncorrected UV LF as follows;
\begin{gather}
    \alpha_{\rm int} = \frac{\alpha-b}{1+b},
    \\
    \phi_{\rm int}^\star = \frac{\phi^\star}{1+b},
    \\
    L_{\rm int}^\star = \eta L^\star = aL^{\star b+1}
\end{gather}
Using these equations, we examine the upper limit for the dust-corrected UV LF by changing $a$ and $b$ under the relation of equation (\ref{Eqn:Relation_ab}), which is in turn used to obtain the upper limit for the SFRD.
Here, we use the Schechter parameters for the dust-uncorrected UV LF at $z\sim8$ given by \citet{bouwens_uv_2015}.
We obtain the upper limit for the dust-corrected rest UV LF with the parameters of $(a,b)=(1.37\times10^{-17},0.637)$, which gives the limit for the SFRD of $\log_{10}(\rho_{\rm SFR}\ [M_\odot\ \mathrm{yr^{-1}\ Mpc^{-3}}])<-1.04$.
The result is shown in Figure \ref{Fig:Uplim_for_SFRD}(b) and (c).

We can immediately see that the upper limits from the two types of parameterization show similar values.
This suggests that the constraint on the total SFRD from the non-detection does not change significantly with the assumption of functional form of the H$\alpha$ LF.
Comparing these constraints with the estimation by \citet{rowan-robinson_star_2016}, even if the dust-obscured SF significantly dominates at $3\lesssim z \lesssim 6$ and the total SFRD is as high as that at $z\sim2$, the total SFRD must decrease moderately by $z\sim8$.

It is worth noting that the result persists with changing assumptions.
Recent observational studies suggest the double power law (DPL) function describes the (dust-uncorrected) rest UV LF at $z\gtrsim8$ better than Schechter function \citep[e.g.][]{bowler_lack_2020}.
Even when DPL is used instead of Schechter function to parameterize the dust-uncorrected rest UV LF, the upper limit does not change significantly.
Further, even if we fix $b$ instead of fixing the relation of $a$ and $b$\footnote{Assuming there are linear relations between $A_{1600}$ and $\beta$ \citep[e.g.,][]{meurer_dust_1999}, and between $\beta$ and $M_{\rm UV}$ \citep[e.g.,][]{bouwens_uv-continuum_2014}, the parameter $b$ can be analytically determined.}, the result is still similar.

\section{Discussion}\label{Sec:Discussion}
\subsection{Effect of Rest UV Selection}
In this study, we used the Lyman Break method to make a sample of galaxies at the target redshift, so the rest UV selection can lead to an incompleteness of our sample and have an effect on the result.
However, the observation in $J_{125}$ and $J\!H_{140}$ band by {\it HST} which we used for the source extraction is much deeper than that in IRAC 5.8 $\mu$m band which we used to measure the H$\alpha$ flux.
Thus, the incompleteness originated from the rest UV selection is expected to have a negligible effect on the result.

\begin{figure}[tpb]
\plotone{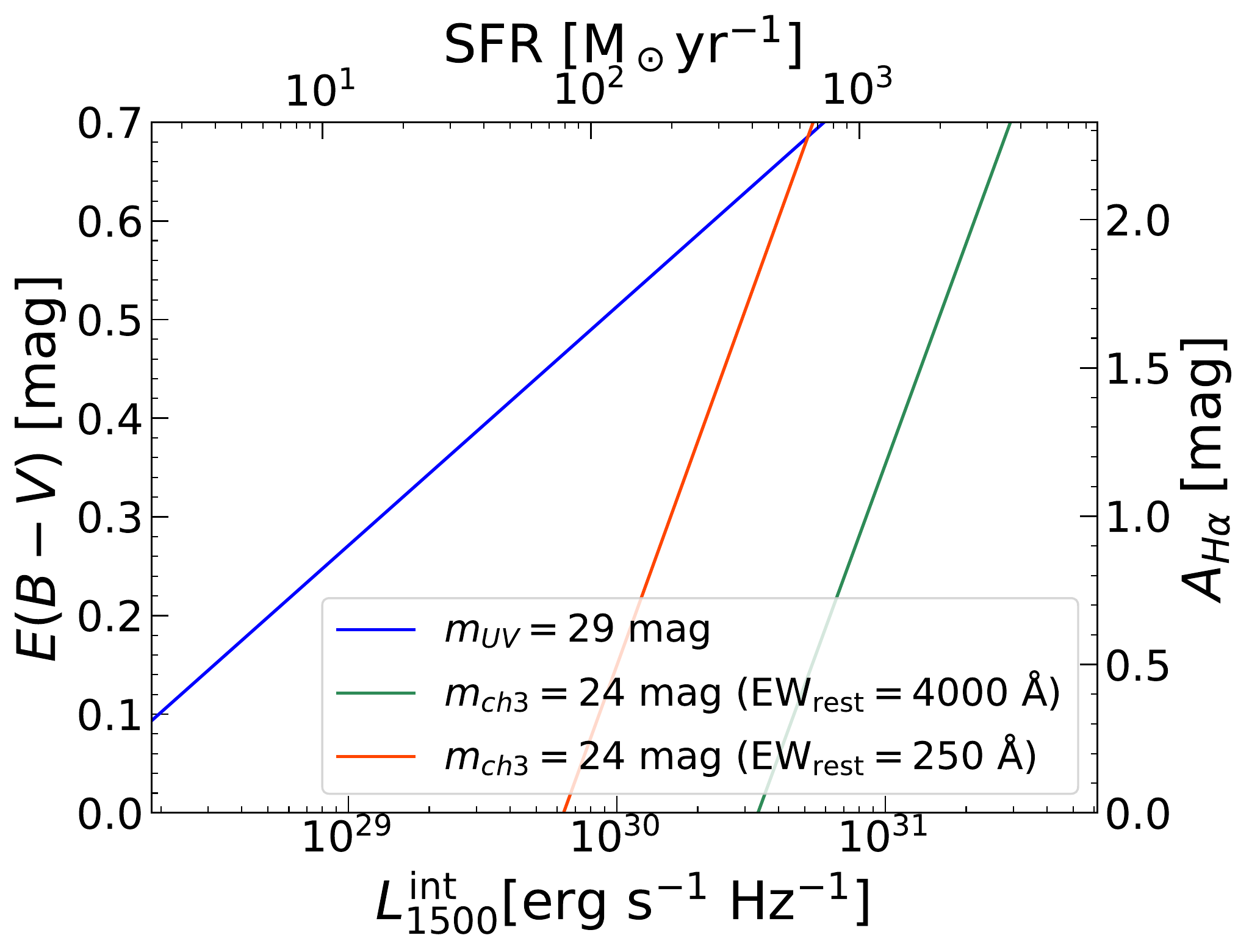}
\plotone{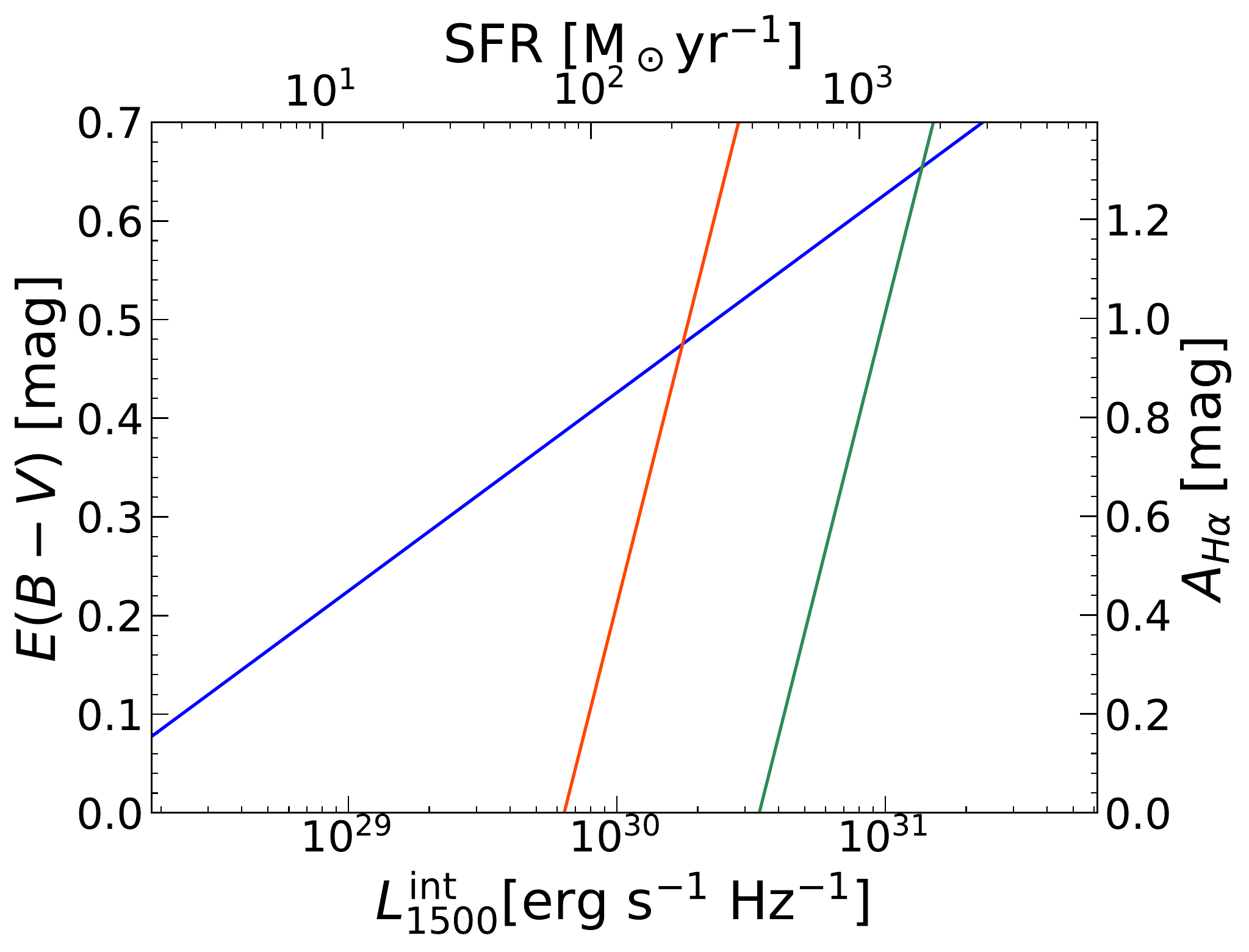}
\caption{Top: Curves of limiting magnitudes of the rest UV observation with {\it HST} and that with {\it Spitzer} in the parameter space of $L_{1500}^{\rm int}$ vs. $E(B-V)$.
Blue line shows the curve that corresponds to the apparent magnitude of $m_{\rm UV}=29$ mag at rest UV (which is the typical limiting magnitude at $J_{125}$ or $J\!H_{140}$ band) in this parameter space.
Red (green) line shows that corresponds to the apparent magnitude of $m_{\rm Ch3} = 24$ mag at IRAC 5.8 $\mu$m band with assuming the H$\alpha$ EW to be $250\ {\rm \AA}$ ($4000\ {\rm \AA}$).
Dust extinction is modeled with Calzetti law \citep[][]{calzetti_dust_2000}.
Bottom: Same as the top panel but for SMC extinction law \citep[][]{pei_interstellar_1992}.}
\label{Fig:Effect_of_Sampling}
\end{figure}

To quantify this, in the parameter space of the (dust-corrected) rest UV luminosity $L^{\rm int}_{1500}$ versus the color excess $E(B-V)$, we compare the region that can be detected with {\it HST} and that can be detected with IRAC 5.8 $\mu$m band.
We assume the redshift of the source to be $z=7.8$, and use equation (\ref{Eqn:UVtoSFR}) and (\ref{Eqn:HatoSFR}) to convert the rest UV luminosity to H$\alpha$ luminosity. 
We also take into account for the contribution by the continuum assuming a value for the rest-frame EW of the H$\alpha$ emission line, and calculate the IRAC 5.8 $\mu$m band magnitude for a given rest UV luminosity.

To determine what value to assume for the EW, we roughly estimate the range of the EW for galaxies that are expected to be detected by this survey.
First, we set the upper limit for the EW to be $4000\ {\rm \AA}$ \citep[e.g.,][]{inoue_rest-frame_2011}.
There is an anti-correlation between the stellar mass and the EW of the H$\alpha$ emission among star-forming galaxies \citep[e.g.,][]{reddy_mosdef_2018}, thus we can estimate the minimum EW that are expected to be detected by estimating the maximum stellar mass.
The effective volume surveyed in this study is $\sim10^4\ {\rm Mpc^3}$.
Considering the galaxy stellar mass function at $z\sim8$ \citep[][]{song_evolution_2016}, the expected value for detecting a galaxy with a stellar mass of $M_\star>10^9\ M_\odot$ is below unity.
Thus, the stellar mass of a galaxy that is expected to be found in this survey is no larger than $10^9\ M_\odot$, which corresponds to the H$\alpha$ EW of $250\ {\rm \AA}$ \citep[][]{reddy_mosdef_2018}\footnote{The relation between the stellar mass and the H$\alpha$ EW by \citet{reddy_mosdef_2018} is derived for star-forming galaxies at $z\sim2.3$, and the redshift is largely different from that in this study. However, this relation has not been probed above $z\sim2.3$, thus we use the relation as the fiducial one.}.
As a result, the expected range of the H$\alpha$ EW is estimated to be $250\ {\rm \AA} \lesssim {\rm EW} \lesssim 4000\ {\rm \AA}$.

The result of comparison is shown in Figure \ref{Fig:Effect_of_Sampling}.
When we make a sample of galaxies based on the rest UV observation with {\it HST}, galaxies locate below the blue line in this figure will be selected.
On the other hand, the constraint obtained in this study is derived from the lack of galaxies below the red (or green) line in this figure.
Thus, as far as we consider the amount of dust extinction on H$\alpha$ emission line up to $\sim2\ {\rm mag}$ ($\sim1\ {\rm mag}$) with Calzetti (SMC) dust extinction law, the effect of missing galaxies due to the rest UV selection on our result is expected to be negligible.

\subsection{The Additional LBG Criteria}\label{subsec:Additional_cri}
In Section \ref{subsec:sample}, we use an additional criteria (equations (\ref{eqn:7}) and (\ref{eqn:8})) to complement our sample with galaxies at $z\sim7$.
In this subsection, we show how the redshift dependence of the selection efficiency by the criteria is evaluated, and discuss about the properties of the sample used in this study.

We first assume the rest UV apparent magnitude and the spectrum of the galaxy (Im-type spectrum by \citet{coleman_colors_1980}), and calculate the position in the color-color diagram for each redshift.
Intergalactic attenuation by neutral hydrogen is modeled following the prescription by \citet{madau_radiative_1995}.
For each position in the diagram, we then mock the observation taking into account for the photometric errors (Gaussian error) for 10000 times, and obtain the 10000 mock colors of the galaxy to calculate the fraction of the mock colors that meet the criteria we used in Section \ref{subsec:sample}.
We examine this fractions for several values of rest UV apparent magnitude, and obtain the effective selection efficiency by taking the weighted mean of them.
Here, we adopt the weight according to the apparent magnitude distribution of galaxies in our sample, resulting in an effective rest UV magnitude of $\sim27.9$ mag.

The result is shown in Figure \ref{Fig:Selection_eff}.
For comparison, the redshift dependence of the selection efficiency by $i_{814}$- and $Y_{105}$-dropout selection are also examined by the same way and shown in this figure.
We can see that the additional criteria cover the redshift range where both of the $i_{814}$- and $Y_{105}$-dropout selection are insensitive.

It is worth noting that we do not exclude low-$z$ interlopers or active galactic nuclei (AGNs) from our sample of galaxies.
Particularly, the additional criteria can pass low-$z$ interlopers, because we do not set the criteria to exclude local brown dwarfs.
However, in this work, we make the sample contain all the galaxies at the target redshifts, even though the sample can be contaminated with low-$z$ interlopers or AGNs, and make the photometry in the IRAC 5.8 $\mu$m band for each galaxy in the sample.
Because no sample galaxy is detected in 5.8 $\mu$m band, the contamination with low-$z$ interlopers or AGNs does not affect on the result.

Moreover, in calculating the constraint on the H$\alpha$ LF, we do not take into account for the effect of the selection efficiency (c.f., equations (\ref{Eqn:13}) and (\ref{Eqn:14})), and we assume the selection efficiency to be unity for galaxies whose redshift is in the target range ($6.9<z<8.6$).
As shown in Figure \ref{Fig:Selection_eff}, the selection efficiency is less unity, thus taking the effect into account can lead to decreasing the effective volume and the forbidden region moves up as a whole.
However, the shift is only slight (less than 0.1 dex), and the upper limit for the SFRD gets larger only by $<0.1$ dex.

\subsection{Other possible uncertainties on the result}
In IRAC photometry, we do not conduct any subtraction of foreground sources.
If an H$\alpha$ emitter at $z\sim7.8$ is blended with a foreground source and obscured in 5.8 $\mu$m band image, such an H$\alpha$ emission can be missed, which leads to decreasing the effective volume of this survey.
However, the effect of foreground sources is expected to be negligible.
Among the 123 LBG candidates, only a few galaxies are potentially blended with a foreground source, and none of them are heavily blended and obscured.
This indicates that the effect of blending decreases the effective volume of this survey only slightly ($\lesssim 2\%$).
The reason for such a small effect is mainly due to the very lower depth in the 5.8 $\mu$m band image.

The presence of foreground sources can also have some effect on LBG detection.
If a low-$z$ source lies in the projection of a $z\sim7.8$ LBG, it prevents us from identifying the LBG.
Such projection effect can also reduces the effective volume of this survey.

To quantify this effect, we place 10,000 apertures randomly and examine the fraction of those meet the shortward wavelength detection criteria ($\textrm{S/N}>2$ in both the $B_{435}$ and $V_{606}$ band or in $B_{435}+V_{606}$ stacked image for $i_{814}$-dropout selection, and $\textrm{S/N}>2$ in $B_{435}$, $V_{606}$, or $i_{814}$ band image for $Y_{105}$-dropout selection).
We obtain the fraction of apertures that meet these criteria corresponds to $\sim6.4\%$ and $\sim17\%$ of the solid angle in the source plane for $i_{814}$-dropout and $Y_{105}$-dropout selection, respectively.
The latter fraction is large to be ignored. However, our sample is complemented with the additional criteria (equations (\ref{eqn:7}) and (\ref{eqn:8})), which use the same criteria for the shortward wavelength detection as $i_{814}$-dropout and covers up to $z\sim8.2$ (Figure \ref{Fig:Selection_eff}).
Therefore, the effect of projection is expected to be small ($\sim10\%$).

It is worth noting that the both effect of blending and projection move the forbidden region not as a whole.
These effects depend on the spatial distribution of the foreground sources, thus the amount of these effects changes depending on the magnification factor and intrinsic H$\alpha$ luminosity (for a fixed observed H$\alpha$ luminosity).

Finally, cosmic variance is also a source of uncertainty on the result.
The area probed by this work is relatively small ($V_{\rm eff}\sim10^4\ {\rm Mpc}^3$), so the uncertainty stems from the cosmic variance may be large.
However, since we find no H$\alpha$ emitter, it is difficult to quantitatively estimate the impact of the cosmic variance.

\subsection{Constraint on the H$\alpha$ EW at $z\sim7.8$}
The forbidden region in Figure \ref{Fig:HaLF_Constraint} and \ref{Fig:Uplim_for_SFRD} depends on the assumed value for the H$\alpha$ EW: if a larger value is assumed for the H$\alpha$ EW, the forbidden region moves right as a whole.
This is because, the larger H$\alpha$ EW we assume, the larger flux in 5.8 $\mu$m band attributes to the H$\alpha$ emission.
Thus, for a fixed limiting flux in 5.8 $\mu$m band observation, the limiting flux of H$\alpha$ emission gets larger (smaller) if a larger (smaller) value is assumed for H$\alpha$ EW.
In this study, we take the maximum value to estimate the most conservative constraint on the H$\alpha$ LF.

On the contrary, the minimum value for the H$\alpha$ EW can be evaluated by changing the assumed value.
Given that the H$\alpha$ LF converted from the (dust-uncorrected) UV LF (blue solid line in Figure \ref{Fig:HaLF_Constraint}) is the lower limit for the H$\alpha$ LF, the forbidden region must not violate this H$\alpha$ LF.
We examine how small the H$\alpha$ EW can be while the forbidden region does not exclude this H$\alpha$ LF.
The minimum value of the H$\alpha$ EW is obtained to be $\sim60\ {\rm \AA}$.
This indicates that the H$\alpha$ EW of the SF galaxies at $z\sim7.8$ is typically larger than $\sim60\ {\rm \AA}$.

\section{Summary}\label{sec:summary}
The amount of the contribution by dust-obscured SF to the total SF activity in high-$z$ universe is still controversial.
The SFRD in high-$z$ universe is mainly probed through rest UV or FIR observations, thus using the H$\alpha$ emission line can give an independent insight.
In this study, we search for H$\alpha$ emitters at $z\sim7.8$ in several gravitationally lensed fields observed in the HFF program.
We make a sample of galaxies at the target redshift with the Lyman break method, and use IRAC 5.8 $\mu$m band to detect H$\alpha$ emissions from galaxies in the sample.
Our main results are as follows:
\begin{enumerate}
    \item We find no significant detection of counterparts in 5.8 $\mu$m band.
    This non-detection gives a constraint on the H$\alpha$ LF at $z\sim7.8$.
    We compare this constraint with previous studies on rest UV LFs at $z\sim8$ and FIR LFs at $z\sim6$ through the SFR.
    The dust-corrected rest UV LF is consistent with the constraint.
    The FIR LF at $z\sim6$ is not consistent with the constraint if we assume the FIR LF does not evolve from $z\sim6$ to $z\sim7.8$.
    Even if we assume the FIR LF evolves from $z\sim6$ to $z\sim7.8$, the FIR LF at $z\sim7.8$ should not violate the constraint, and this may suggest a negative evolution of the FIR LF from $z\sim6$ to $z\sim7.8$ (Section \ref{subsec:constraint_HaLF} and Figure \ref{Fig:HaLF_Constraint}).
    \item We put a constraint on the SFRD at $z\sim7.8$ assuming the shape of H$\alpha$ LF.
    We examine two types of functional form, and obtain an upper limit for the SFRD as $\log_{10}(\rho_{\rm SFR}\ [M_\odot\ \mathrm{yr^{-1}\ Mpc^{-3}}])\lesssim -1.1$ (Section \ref{subsec:constraint_SFRD} and Figure \ref{Fig:Uplim_for_SFRD}).
    \item With the constraint on the SFRD at $z\sim7.8$, even if the dust-obscured SF dominates significantly at $3\lesssim z \lesssim 6$ and the total SFRD at this epoch is as large as that at $z\sim2$-3 \citep[e.g.,][]{rowan-robinson_star_2016}, the total SFRD must decrease moderately by $z\sim8$ (Section \ref{subsec:constraint_SFRD} and Figure \ref{Fig:Uplim_for_SFRD}).
\end{enumerate}

We thank the anonymous referee for useful suggestions and comments.
We also thank Fumiya MAEDA and Marcin SAWICKI for meaningful discussions to improve this paper.
KO is supported by JSPS KAKENHI Grant Number JP19K03928.
This work is based on observations obtained with the NASA/ESA Hubble Space Telescope, retrieved from the Mikulski Archive for Space Telescopes (MAST) at the STScI. STScI is operated by the Association of Universities for Research in Astronomy, Inc. under NASA contract NAS 5-26555.
This work utilizes gravitational lensing models produced by PIs Natarajan \& Kneib (CATS).
This lens modeling was partially funded by the HST Frontier Fields program conducted by STScI.
STScI is operated by the Association of Universities for Research in Astronomy, Inc. under NASA contract NAS 5-26555.
The lens models were obtained from the MAST.

%If $L^{\rm int}_{H\alpha}$ is larger than the apparent limiting luminosity, such a galaxy can be detected with this observation wherever it locates, thus the effective volume

\software{{\sc SExtractor} \citep[][]{bertin_sextractor_1996}, IRAF \citep[][]{1986SPIE..627..733T,1993ASPC...52..173T}, Astropy \citep[][]{the_astropy_collaboration_astropy_2018}, APLpy \citep[][]{2012ascl.soft08017R}}


\bibliography{my_library}{}
\bibliographystyle{aasjournal}



\end{document}